\documentclass[aps, pre, reprint]{revtex4-1}

\usepackage[utf8]{inputenc} 
\usepackage[T1]{fontenc}    
\usepackage{hyperref}       
\usepackage{url}            
\usepackage{booktabs}       
\usepackage{amsfonts}       
\usepackage{nicefrac}       
\usepackage{microtype}      
\usepackage{xcolor}         
\usepackage{bm}
\usepackage{amsmath}
\usepackage{amsthm}
\usepackage{graphicx}
\usepackage{mathtools}
\usepackage{amsmath}
\usepackage{graphicx}
\newtheorem{definition}{Definition}
\usepackage{algorithm}
\usepackage{algorithmic}
\usepackage{bbm}

\begin{document}


\title{Frustrated Random Walks: A Fast Method to Compute Node Distances on Hypergraphs}


\author{Enzhi Li}
\email{lienzhi@amazon.com}
\affiliation{Amazon, San Diego, USA}
\author{Scott Nickleach}
\email{nickleac@amazon.com}
\author{Bilal Fadlallah}
\email{bhf@amazon.com}
\affiliation{Amazon, Seattle, USA}


\date{\today}

\begin{abstract}
A hypergraph is a generalization of a graph that arises naturally when attribute-sharing among entities is considered. Compared to graphs, hypergraphs have the distinct advantage that they contain explicit communities and are more convenient to manipulate. An open problem in hypergraph research is how to accurately and efficiently calculate node distances on hypergraphs. Estimating node distances enables us to find a node's nearest neighbors, which has important applications in such areas as recommender system, targeted advertising, etc. In this paper, we propose using expected hitting times of random walks to compute hypergraph node distances. We note that simple random walks (SRW) cannot accurately compute node distances on highly complex real-world hypergraphs, which motivates us to introduce frustrated random walks (FRW) for this task. We further benchmark our method against DeepWalk, and show that while the latter can achieve comparable results, FRW has a distinct computational advantage in cases where the number of targets is fairly small. For such cases, we show that FRW runs in significantly shorter time than DeepWalk. Finally, we analyze the time complexity of our method, and show that for large and sparse hypergraphs, the complexity is approximately linear, rendering it superior to the DeepWalk alternative.
\end{abstract}


\maketitle

\section{Introduction}
\begin{figure}[h!]
\centering
\includegraphics[width=0.85\linewidth]{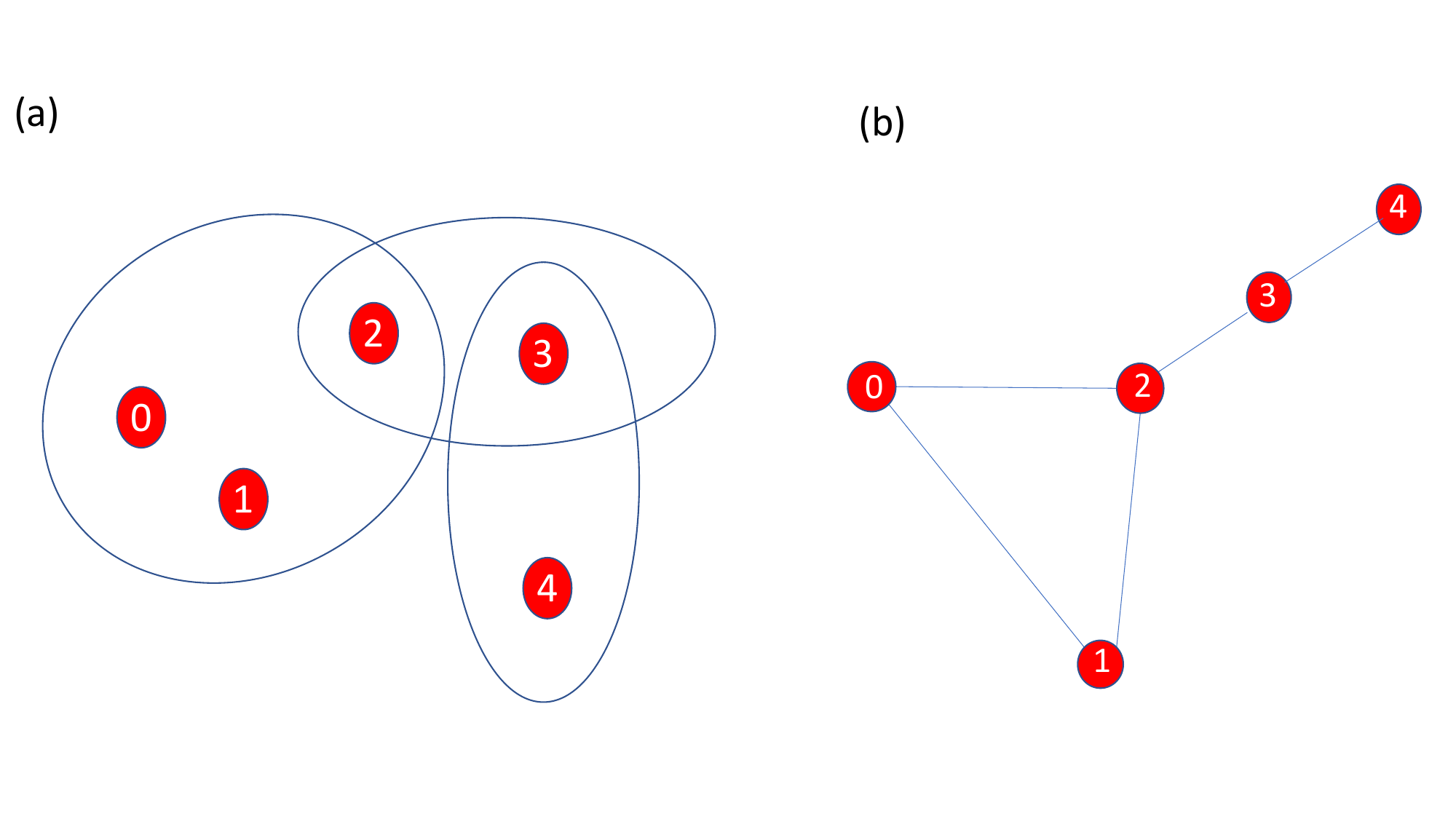}
\caption{Panel (a): A hypergraph; panel (b): The expanded graph. The conversion of graph to hypergraph is possible in theory, yet almost impossible in practice. }
\label{graph_hypergraph}
\end{figure}

A graph is a useful data structure for describing complex relations in real world, such as users on social media, audiences of movies, etc\cite{newman2018networks}. In a graph, nodes are connected by edges, and each edge connects exactly two nodes. A hypergraph is an extension of ordinary graphs where each hyperedge of a hypergraph may contain an arbitrary number of nodes\cite{zhou2006learning}. In the special case when each hyperedge contains just two nodes, the hypergraph reduces to a graph. Each hyperedge of a hypergraph is considered a clique in a graph. We can therefore convert a hypergraph into a graph by expanding each hyperedge. Fig. \ref{graph_hypergraph} illustrates a hypergraph and its corresponding expanded graph. 

Hypergraphs arise naturally when there are explicit communities of nodes. In social media, for instance, users are represented as nodes, and users belonging to the same group are collected into the same hyperedge. With the hypergraph representation of users, we want to quantify the similarity between any pair of users, and use this information to suggest friends or recommend products. Intuitively, the similarity between users should be negatively correlated with their distance, thus the computation of hypergraph node distances is a prerequisite for the implementation of recommender systems and targeted advertising\cite{sarwar2000application, fouss2007random, wu2022graph}. 

Although we can easily convert a hypergraph into a graph by expanding its hyperedges, we claim that hypergraphs contain more information than what can be encapsulated into its expanded graphs due to the explicit presence of communities in the former. Furthermore, the expansion of hyperedges can be computationally expensive and memory-intensive, thus making it advantageous to directly work with hypergraphs whenever possible rather than with their expanded graphs.

More formally, a hypergraph consists of nodes and hyperedges. Each hyperedge is a subset of the node set $\mathbb{V}$. For clarity purpose, we use in this paper Latin letters to indicate node indices and Greek letters to indicate hyperedge indices. The incidence matrix of a hypergraph is defined as: 
\begin{align}
e_{i\alpha} = \begin{cases}
& w, \text{if vertex } v_{i} \in E_{\alpha} \\
& 0, \text{otherwise}
\end{cases}
\end{align}

In the above definition, $w > 0$ is the weight of vertex $v_i$ in hyperedge $E_{\alpha}$. If the hypergraph is unweighted, then $w = 1$ always holds. If we think of $v_{i}$ as a member of a community $E_{\alpha}$, then $e_{i\alpha}$ can be thought of as the loyalty of $v_{i}$ to $E_{\alpha}$. By definition, the degree of a node $v_{i}$ is $D_{i} = \sum_{\alpha} e_{i\alpha}$, and the degree of a hyperedge is $\delta_{\alpha} = \sum_{i} e_{i\alpha}$. We can think of $\delta_{\alpha}$ as the adhesiveness of $E_{\alpha}$. 

Given the important applications of computing hypergraph node distances, we have developed a method based on random walks for this task. We will use the expected hitting times of random walks as node distances on hypergraphs. We will show that the simple random walk method is insufficient for the description of complex human-human interactions, and propose our \textit{frustrated random walk} method. In frustrated random wallks, we augment each transition with an acceptance probability, and thus obtain a different transition matrix from the one in simple random walks. Although the simple and frustrated random walks have different transition matrices, we have developed a unified theoretical framework to efficiently compute their expected hitting times. 

In summary, we made three major contributions in our work. First, we propose frustrated random walks on hypergraphs, a random walk scenario that outperforms simple random walks in computing node distances for heavily-weighted and scale-free hypergraphs. Second, we design a unified theoretical framework for computing the expected hitting times of random walks, whether it be simple or frustrated, and use the expected hitting times as hypergraph node distances. Third, we test our methods on real-world hypergraph datasets and demonstrate that frustrated random walk method significantly outperforms existing methods in accuracy and/or running speed.

\section{Background}
The importance of the computation of graph node distances has already been widely acknowledged\cite{sommer2016comparison}, with
significant applications in social networks (friend suggestion, link prediction, targeted advertising), e-commerce (product recommendation), etc. Intuitively, the shorter the distance between two nodes, the more similar they are to each other. Therefore, the primary purpose of computing node distances is to determine node similarities, thus enabling the identification of the nearest neighbors of a target node within a graph. An obvious candidate for computing graph node distances is Dijkstra's algorithm, which can quickly provide the shortest or geodesic distance from a source node to a target node. Although this algorithm is quick, it focuses solely on the shortest path and ignores all possible alternative routes. However, in a graph representation of social networks, the similarity between two nodes is expected to increase with the number of available alternative routes connecting them. This expectation has led to the development of algorithms that consider the entire graph structure when computing node distances. A plethora of such algorithms already exist in the literature, which can be crudely classified into two categories.  Below, we provide a brief review of these algorithms, focusing only on those most relevant to our work, rather than attempting an exhaustive review of all existing methods.

The first category of algorithms relies on the study of random walks on graphs, and uses expected hitting time (also known as first-passage time) and commute time to evaluate node distances\cite{klein1993resistance, white2003algorithms, fouss2007random, PhysRevE.102.052135}. In this framework, for a pair of graph nodes denoted as $v_i$ and $v_j$, the hitting time $N_{j}^{(i)}$ of a random walk process is defined as the number of steps a random walker starting from node $v_i$ needs to traverse before it hits node $v_j$ for the first time. $N_{j}^{(i)}$ is a random variable whose expectation value $\mathbb{E}N_{j}^{(i)}$ can be used to measure the distance from $v_i$ to $v_j$. Here, the distance should be understood as the dissimilarity between nodes $v_i$ and $v_j$ since it does not satisfy the symmetricity and triangle-inequality conditions that are required of a mathematical distance function. From the expected hitting times, we can construct the expected commute time, defined as $n(i, j) = \mathbb{E}N_{j}^{(i)} + \mathbb{E}N_{i}^{(j)}$, which satisfies all the conditions of a distance function in mathematics, and is therefore a genuine distance in the strict mathematical sense. Given a graph structure, the closed form expressions for $\mathbb{E}N_{j}^{(i)}$ and $n(i, j)$ can be readily obtained from the pseudo-inverse of the graph Laplacian, as is shown in Ref. \cite{fouss2007random}. It is however shown in Ref. \cite{PhysRevE.102.052135} that for heavily-weighted and scale-free graphs whose node degree distribution follows a power law, the expected hitting times of simple random walks cannot accurately evaluate node distances, which necessitates the introduction of frustrated random walks to graphs. Here, we will generalize these methods to hypergraphs, and show that frustrated random walks have distinct advantages in measuring node distances of such highly complicated hypergraphs. 

The second category of algorithms relies on graph node embedding, also called graph representation learning, which means mapping graph nodes to low-dimensional dense vectors and using the cosine distances between these vectors as graph node distances\cite{hamilton2017representation}. This includes such algorithms as DeepWalk\cite{perozzi2014deepwalk}, node2vec\cite{grover2016node2vec}, structural deep network embedding (SDNE)\cite{wang2016structural}, graphSAGE\cite{hamilton2017inductive}, graph attention networks\cite{velivckovic2017graph}, etc. These algorithms aim to map graph nodes to a continuous vector space so that nodes that are neighbors in the original graph are mapped to vectors with small cosine distances. There is no universal definition of neighborhood in these algorithms. For example, in DeepWalk, node2vec and graphSAGE, nodes that frequently co-appear within a pre-specified window in sampled random paths are considered neighbors, whereas in SDNE, neighboring nodes are those that are of first-order or second-order proximity with respect to each other in the original graph. Despite these differences, all algorithms adhere to the guiding principle that neighboring nodes should be mapped to neighboring vectors during the node embedding process. After node embedding, we can use cosine distances between mapped node vectors as graph node distances. Generally, the cosine distances from the node embedding methods is a good metric for evaluating graph node distances, however complex the graphs may be. Among these algorithms, the DeepWalk method excels at measuring node distances of scale-free graphs. A second advantage of DeepWalk in comparison with the other node embedding algorithms is the ease with which it can be generalized to hypergraphs, which prompts us to use it as our benchmark method in our work. 

With the abundance of algorithms for computing graph node distances, a natural next step is to generalize them to hypergraphs. The generalization of DeepWalk from graphs to hypergraphs is trivial. To understand why, a brief review of DeepWalk method is necessary. To accomplish graph node embedding using DeepWalk, we first perform random walks on a graph, then record the paths that were generated during the random walk process, interpret the nodes as words and the random walk paths as corpus, and finally invoke word2vec\cite{mikolov2013efficient, mikolov2013distributed} to map graph nodes to vectors. Since the generation of random paths in this method is agnostic about whether we are performing random walks on graphs or hypergraphs, a generalization of it to hypergraphs is straightforward. 

Although we can easily generalize DeepWalk to hypergraphs, for the other algorithms, such a generalization is highly non-trivial. In this paper, we generalize frustrated random walks to hypergraphs, and show that its performance is on par with the that of DeepWalk even for highly complicated hypergraphs. Compared to DeepWalk, frustrated random walks (FRW) has the following four advantages. First, in applications where the goal is to find the nearest neighbors of a few nodes in a large hypergraph, FRW is a preferable and faster option than DeepWalk. Second, FRW conveniently gives a closed-form and interpretable solution for node distances, a solution that is impossible to obtain using traditional deep-learning-based methods. Third, FRW does not require any parameter tuning, whereas DeepWalk requires heavy investment in this task. Fourth, and unlike that in DeepWalk or any other node embedding algorithm, the node distances of FRW are asymmetric, rendering it more suitable to describe real-world relationships which are generally non-equivalent and non-reflective.

Researchers have long used random walks to study hypergraphs. In Ref. \cite{zhou2006learning}, the authors generalized spectral clustering  \cite{shi2000normalized, meilua2000learning, ng2002spectral} from graphs to hypergraphs and gave their algorithm a random walk interpretation. It is however noted that the hypergraph random walks defined in Ref. \cite{zhou2006learning} are no different from the random walks performed on expanded graphs \cite{agarwal2006higher, chitra2019random}. To take advantage of the higher order structure in hypergraphs, we need to take hyperedge degree into account when performing random walks. Since a hyperedge in a hypergraph represents an adhesive community, it is argued in Ref. \cite{PhysRevE.101.022308} that a random walker roaming on a hypergraph should show preference towards hyperedges of higher degrees or stronger adhesiveness. This is not the only way to generalize random walks from graphs to hypergraphs, but it does demonstrate its advantages, as shown in Ref. \cite{PhysRevE.101.022308}. We also note that in Ref. \cite{PhysRevE.101.022308}, the authors assume that all the hypergraph nodes have the same weight, thus restricting their results only to unweighted hypergraphs. Here in this paper, we will generalize the random walks described in Ref. \cite{PhysRevE.101.022308} to weighted hypergraphs, and introduce the concept of frustrated random walks on hypergraphs. We will show that for heavily-weighted and scale-free hypergraphs, the exact definition of which will be given later, frustrated random walks are more suitable for computing node distances. 

Without loss of generality, we will assume throughout the paper that the hypergraphs are connected, meaning that there exists a path between any pair of nodes on a hypergraph. The ease with which we can detect connected components of a hypergraph ensures that this assumption does not compromise the applicability of our method, since we can always apply our method to each connected component of a disconnected hypergraph. 

\section{A unified framework for calculating expected hitting times of random walks}
Despite the cornucopia of research on random walks on hypergraphs, the vast majority of previous work has focused on the Laplacian matrix. Spectral clustering of eigenvectors of Laplacian matrices \cite{shi2000normalized, ng2002spectral} naturally leads to image segmentation and community detection in hypergraphs \cite{hayashi2020hypergraph, carletti2021random}. Furthermore, the eigenvectors of a Laplacian matrix yield node embeddings which enable us to apply powerful machine learning algorithms to perform node classification on hypergraphs. Since the study of the Laplacian matrix yields information about the long-term stationary distribution of particles that are randomly roaming on hypergraphs, we call this method of studying random walks the stationary approach. On the other hand, we can also extract precious information about hypergraphs by focusing on the diffusion process itself, and refer to this approach to random walks as being dynamic. 

In this paper, we will adopt the dynamic approach by focusing on the hitting times of random walks. Given a hypergraph $H$, we select a node $t$ as target, and perform random walks starting from any other node $s$. The \textit{hitting time} for this process is defined as the number of steps a random walker needs to traverse before it hits the target $t$ for the first time. By definition, the hitting time on a hypergraph is a random variable that depends on the hypergraph structure, the starting node $s$, the target node $t$, and can thus be denoted as $N^{(s)}_t$. The expectation of $N^{(s)}_t$ provides a natural measurement of the distance from $s$ to $t$. Note that the distance here is asymmetric, meaning that the distance from $s$ to $t$ is not guaranteed to be identical to that from $t$ to $s$. Since we want to use the expected hitting time to measure the closeness between two persons in the real world, and we know that the real-world human-human relationships are generally asymmetric, this lack of symmetry is thus considered desirable. 

The dynamics of random walks on hypergraphs are fully captured by the transition matrix, and different definitions of transition matrices produce different expected hitting times. We calculate hypergraph node distances to find the nearest neighbors of a node. To achieve this, we select a node as the target, and calculate the expected hitting times starting from all other nodes. We then rank all those nodes according to their distances with respect to the target, and select the top-$N$ nodes as the target's nearest neighbors. If we represent real-world human-human interactions using a hypergraph, we expect to find a person's close friends from among his/her nearest neighbors in the hypergraph. In this paper, we will evaluate the quality of our method's results by creating a hypergraph representation of real-world human relationships and checking whether a person's close friends show up in his/her top neighbor list. 

\subsection{Simple random walks on hypergraphs}
\label{SRW_section}
As already stated, the computation of hypergraph hitting times requires specification of the random walk's transition matrix. To capture a random walker's preference towards a more cohesive hyperedge, we generalize the method in Ref. \cite{PhysRevE.101.022308} from unweighted to weighted hypergraphs, and define transition probability from node $i$ to node $j$ as 
\begin{align}
T^{S}_{ij, i\ne j} = \frac{\sum_{\alpha}\Big( \delta_{\alpha} - e_{i\alpha}\Big) \min\{e_{i\alpha}, e_{j\alpha}\}}{\sum_{k \ne i }\sum_{\alpha}\Big( \delta_{\alpha} - e_{i\alpha}\Big) \min\{e_{i\alpha}, e_{k \alpha}\}}
\label{srw_transition_matrix_new}
\end{align}
The mechanism behind Eq. (\ref{srw_transition_matrix_new}) is that a random walker standing on site $i$ would randomly select one of its neighbors $j$ towards which it makes a transition. A transition from $i$ to $j$ is possible only when both nodes belong to the same hyperedge. In a hypergraph that represents real-world human-human interactions, the node weight $e_{i\alpha}$ measures a member's loyalty to the hyperedge which is interpreted as a community. Intuitively, two loyal members tend to have high frequency interactions, and two disloyal members tend not to interact with each other. The interaction frequency between a loyal member say $i$ and a disloyal member say $j$ depends on how disloyal $j$ is to the community. Thus, when calculating the transition probability from $i$ to $j$, we demand that the probability should be proportional to the lesser of the two node weights, i.e., $ \min\{e_{i\alpha}, e_{j\alpha}\}$, rather than their product $e_{i\alpha} e_{j\alpha}$. The proportionality coefficient $\delta_{\alpha} - e_{i\alpha}$ in front of $ \min\{e_{i\alpha}, e_{j\alpha}\}$ reflects the tendency of a node to stay within a more cohesive community (a hyperedge of higher degree after subtracting node $i$'s contribution). The denominator in Eq. (\ref{srw_transition_matrix_new}) is the normalization constant.  

\subsection{Insufficiencies of simple random walks}
\label{srw_insufficiency}
Although simple random walk as described in Section \ref{SRW_section} is a good foundational algorithm for exploring hypergraph properties and can be used to generate random paths for DeepWalk, we find that the expected hitting times derived from this algorithm are not a good measurement of node distances when the hypergraph is heavily weighted and scale-free\cite{barabasi1999emergence}. Using a hypergraph representation of characters in \textit{Harry Potter} novels, the details of which data set will be given in Section \ref{scale_free_heavily_weighted}, we illustrate the limitations of simple random walks. 

In this hypergraph data set, Harry Potter is the central node with the highest node degree, and we want to find his close friends in this hypergraph. From our knowledge of the novels, we know that Ron Weasley and Hermione Granger are his close friends. Thus, if the expected hitting times of simple random walks is a good measurement of node distances, then Ron Weasley and Hermione Granger should have the smallest distances with respect to Harry Potter. Yet, after computing the expected hitting times of simple random walks with Harry Potter as the target node, we find that Hokey who is a house-elf has the smallest distance. The reason is that Hokey appears in the novels only once, and this single appearance coincides with that of Harry Potter. Consequently, in the hypergraph, Hokey has only a few connections, and all of these connections are directed towards Harry Potter. Therefore, a random walker starting from Hokey can rapidly hit Harry Potter. On the other hand, Hermione Granger, who is the heroine of the novels, has many connections with other characters than Harry Potter, and thus the expected hitting time from her to Harry Potter would be larger. 

This observation compels us to conclude that simple random walks are insufficient for measuring node distances in highly complex hypergraphs, particularly concerning the protagonists. To accurately measure node distances, we need to design a mechanism to facilitate the transition from one protagonist to another while frustrating the transitions from minions to protagonists. This led to our proposal of the concept of frustrated random walks, which we will elaborate in the next section. 

\subsection{Frustrated random walks on hypergraphs}
In the previous section, we demonstrated the insufficiencies of simple random walks using a hypergraph constructed from the \textit{Harry Potter} novels. A key characteristic of this hypergraph is that the protagonists, though much fewer in number compared to the minions, have significantly higher node degrees. Hypergraphs with such a feature are mostly scale-free, meaning that their node degrees follow the power law distribution. Scale-free hypergraphs emerge naturally when generated following the principle of preferential attachment, which means when we introduce a new node to an existing graph or hypergraph, we attach it to a node with probability proportional to the node degree. Graphs or hypergraphs that are generated following this principle are guaranteed to be scale-free\cite{barabasi1999emergence}. 

Preferential attachment is ubiquitous in real world, with examples including the accumulation of personal wealth, the number of times of a scientific paper being cited, the number of followers an account attracts in social media, etc. This is equivalent to the Matthew Effect, which postulates that the more one has, the easier it is for one to acquire even more. For instance, in the \textit{Harry Potter} novels, a random character has a much higher probability of interacting with the protagonists than with the peripheral characters. Similarly, in academia, a highly cited researcher has a much higher probability of being cited by a new paper than an obscure one. Preferential attachment means a central node can easily be linked with lots of peripheral nodes, who would then make rapid transitions to the central node under the scenario of simple random walks. Consequently, when we use the expected hitting times of simple random walks to measure node distances, we will find that the nearest neighbors of a central node are mostly peripheral ones. We want to avoid this by introducing the concept of frustrated random walks. 



To address the unique structure of scale-free hypergraphs, we introduce a frustration mechanism to random walks. This approach aims to suppress transitions from peripheral nodes to central nodes while preserving transitions among central nodes. In measuring node distances on hypergraphs using random walks, we designate one node $t$ as the target and initiate random walks from another node $s$. The expected hitting time of this random walk process is interpreted as the distance from $s$ to $t$. Intuitively, the shorter the distance, the closer the relationship  between $s$ and $t$. To more accurately measure node distances in real-world hypergraphs, we can conceptualize this random walk process as analogous to the donation of a gift. If one person has one gift to donate, she tends to give it to her close friend. Similarly, when a person is offered multiple gifts, she tends to accept her close friend's. With this observation, we decompose the gift-donating process into two parts: one for proposing the donation, and another for accepting it. By introducing the acceptance probability to the gift donation, we have \textit{frustrated} the gift donation process, from whence arises our algorithm's name. Under this scenario, a transition is possible only when a proposal is made and then accepted. Therefore, in frustrated random walks, the transition probability is the product of the proposal probability and the acceptance probability, which is

\begin{align}
\label{frustrated_transition_probability}
T^F_{ij, i\ne j} & = \frac{\sum_{\alpha}\Big( \delta_{\alpha} - e_{i\alpha}\Big) \min\{e_{i\alpha},  e_{j\alpha}\}}{\sum_{k \ne i}\sum_{\alpha}\Big( \delta_{\alpha} - e_{i\alpha}\Big) \min\{e_{i\alpha},  e_{k \alpha}\}} \\\nonumber
&\times
\frac{\sum_{\beta}\Big( \delta_{\beta} - e_{j\beta}\Big)\min\{ e_{j\beta}, e_{i\beta}\}}{\sum_{l \ne j} \sum_{\beta}\Big( \delta_{\beta} - e_{j\beta}\Big)\min\{ e_{j\beta}, e_{l \beta}\}} 
\end{align}

To see the concrete difference between Eq. (\ref{srw_transition_matrix_new}) and Eq. (\ref{frustrated_transition_probability}), we will use the \textit{Harry Potter} data set as an example to calculate the one-step transition probabilities from Hokey and Hermione Granger to Harry Potter using these two equations. The results are shown in Table \ref{harry_potter_transition_probabilities}. From the table, we can see that in simple random walks, the one-step transition probability from Hokey to Harry Potter is 1.7 times as large as that from Hermione Granger to Harry Potter. The situation reverses in frustrated random walks, for which the one-step transition probability from Hermione Granger to Harry Potter is 289 times larger than that from Hokey to Harry Potter. Since we know that Hermione Granger is a close friend of Harry Potter, the result from frustrated random walks is more consistent with our intuition. 

\begin{table*}
\centering
\begin{tabular}{ccc}
\hline\hline
Method name & $T_{\text{Hokey} \rightarrow \text{Harry Potter}}$ & $T_{\text{Hermione Granger}\rightarrow \text{Harry Potter}}$  \\
\hline
Simple random walks & $\frac{19}{72}\approx 0.264$ &  $\frac{9380}{60803}\approx 0.154$  \\
Frustrated random walks & $\frac{19}{72} \times \frac{19}{93678} \approx 5.35\times 10^{-5}$ & $\frac{9380}{60803} \times \frac{9380}{93678} \approx 0.015$ \\
\hline\hline 
\end{tabular}\\
\caption{Comparison of one-step transition probabilities from Hokey and Hermione Granger to Harry Potter, in simple and frustrated random walk scenarios. The fact that the transition probability from Hermione Granger to Harry Potter is almost 300 times larger than that from Hokey to Harry Potter in the frustrated random walks authenticates our algorithm, demonstrating its ability to reflect more intuitive relationships.}
\label{harry_potter_transition_probabilities}
\end{table*}

Another major difference between simple random walks and frustrated random walks is that in the former scenario, the proposal to make a transition is always accepted and thus $T^S_{ii} = 0,  \forall i \in \mathbb{V}$, whereas in the latter scenario, there is a non-zero probability for the proposal to be declined, and thus $T^F_{ii}$ is generally non-zero. The diagonal values of $T^F$ matrix can be easily calculated from the observation that $\sum_{j}T^F_{ij} = 1, \forall i, j \in \mathbb{V}$, from which we have 
\begin{align}
T^F_{ii} = 1- \sum_{j \ne i} T^F_{ij}, \forall i \in \mathbb{V}. 
\end{align}
All of these differences combined impact significantly on the computation of expected hitting times and ultimately the ranking of nodes with respect to a target node. We will show later, using real-world data, that frustrated random walks can better capture human-human relationships in the real world than simple random walks. 

\subsection{A Unified framework for calculating expected hitting times}
\label{unfiedframework}
In the previous sections, we have outlined simple random walks and frustrated random walks on hypergraphs, and formulated equations to calculate their corresponding transition probabilities. In this section, we extend the calculation to derive the expected hitting times from the transition probabilities. We show that although the transition probabilities of simple and frustrated random walks differ, the calculation of expected hitting times can be formulated in a unified framework. 

For a connected hypergraph $H$, denote $N_{t}^{(s)}$ as the hitting time of a random walk process that starts from node $s$, with node $t$ as the target. We use $P(N_{t}^{(s)} = n)$ to denote the probability of reaching target node $t$ from node $s$ after exactly $n$ steps. Since the problem is trivial for the case when $s = t$,  we demand that the starting node $s$ differ from the target node $t$. Due to the Markov property of the random walk process, whether it be simple or frustrated, we can establish a recurrence equation for the probability, which is 
\begin{align}
P(N_{t}^{(s)} = n) = \sum_{i_s \ne t} T_{s, i_s} P(N_{t}^{(i_s)} = n-1), n \ge 2
\label{recurrence_equation}
\end{align}
A physical interpretation of the above equation is that we decompose the random walk process into two steps. First, we make a transition from node $s$ to one of its neighbors $i_{s}$; second, due to the Markov property of the random walks, the whole process restarts all over again with $i_s$ as the new starting node, on the condition that now the random walker needs to reach the target with exactly $n-1$ steps since one step has already been taken. Thus, Eq. (\ref{recurrence_equation}) is a recurrence equation for the probabilities $P(N_{t}^{(s)} = n)$. In this equation, we have imposed the condition that $n \ge 2$ because $P(N_{t}^{(j)} = 0) = 0$ for any node $j \ne t$, and the recursion process terminates immediately when the random walker hits the target. From the transition matrix in Eq. (\ref{srw_transition_matrix_new}) or Eq. (\ref{frustrated_transition_probability}), we can readily read out the one-step transition probability $P(N_{t}^{(s)} = 1)$, which is the initial condition for the recurrence relation in Eq. (\ref{recurrence_equation}). 

Eq. (\ref{recurrence_equation}) is essentially a system of first order difference equations with constant coefficients. Denoting $N_{t}^{(i)}$ as the $i$th component of a column vector $\boldsymbol{N}_{t}$, and $P(N_{t}^{(i)} = n)$ as the $i$th component of a column vector $\boldsymbol{X}_{n} = P(\boldsymbol{N}_{t} = n)$, we can rewrite Eq. (\ref{recurrence_equation}) as 
\begin{align}
\boldsymbol{X}_{n} = B \boldsymbol{X}_{n-1}, n \ge 2. 
\label{recursive_equation}
\end{align}
Here, $B$ is the coefficient matrix with elements depending on the hypergraph structure, the random walk scenario (simple or frustrated random walk), and the target node $t$. See the appendix \ref{appendix_1} for an illustration of how to calculate the $B$ matrix. 

\subsubsection{Calculation of the $B$ matrix for simple random walks}
\label{SRW_B_matrix}
For simple random walks with node $t$ as the target, $B_{ij} = T^S_{ij} > 0, \forall i, j \in \mathbb{V} - \{t\}$, and $B_{it} = 0,  \forall i \in \mathbb{V} - \{t\}$ by definition of Eq. (\ref{recurrence_equation}), where $\mathbb{V}$ is the hypergraph vertex set. This means $B$ is a non-negative matrix. Still from Eq. (\ref{recurrence_equation}), we have that the left subscript of $B$, which is the index of the starting node for the random walk, can never be equal to $t$. In contrast with the transition matrix $T^S$, matrix $B$ is not a Markov matrix because the summation of each row does not always equal to one. This non-Markov property of $B$ originates from the existence of the target node $t$. From Eq. (\ref{recurrence_equation}), we get the rule that
\begin{align}
\sum_{j} B_{ij} = 
\begin{cases}
1 & \text{ if } t \text{ is not adjacent to } i \\
1 - T^S_{it} < 1 & \text{otherwise}
\end{cases}
\end{align}
The connectedness of the hypergraph ensures the irreducibility of matrix $B$. Such a matrix as $B$ is called a substochastic matrix. According to Perron–Frobenius theorem for irreducible and non-negative matrices, the spectral radius of matrix $B$, denoted as $\rho(B)$, satisfies the following inequality\cite{meyer2023matrix}: 
\begin{align}
\label{spectral_radius_range}
\min_{i}\sum_{j}B_{ij} \le \rho(B) \le \max_{i}\sum_{j} B_{ij} = 1
\end{align}
It can further be proved that the spectral radius of matrix $B$ must be smaller than one \cite{666603}. A heuristic argument proof of this theorem is given in Appendix \ref{appendix_2}. We will use this fact to quickly compute expected hitting times in this paper. 

We have established Eq. (\ref{recurrence_equation}) to calculate the probability distribution of hitting times using any non-target node $s$ as our starting node. However, in the special case where the starting node $s$ has only one neighbor, and this very neighbor is precisely the target node $t$, then no recurrence equation is needed since we already have $P(N_{t}^{(s)} = n) = \delta_{n, 1}, \forall n \ge 1$. We call such a node an \textit{adherent} node of target node $t$, and exclude these nodes from the set of starting nodes in Eq. (\ref{recurrence_equation}) (For example, node 4 in Fig. \ref{hypergraphexample} is an adherent to target node 3.). 
\begin{definition}
  In random walks, a node is called an adherent to the target if the node has the target as its only neighbor. 
\end{definition}
Since we need to exclude target node $t$ and its adherent nodes from the starting node set, the dimension of matrix $B$ is 
\begin{align}
N_{B} = |\mathbb{V} - \{t\}| - |\{v_{i}: T^S_{it} = 1\}|, 
\end{align}
where $|\mathbb{S}|$ indicates the cardinality of a set $\mathbb{S}$. 
Obviously, for simple random walks, the hitting time from an adherent to the target is always equal to one. 

\subsubsection{Calculation of the $B$ matrix for frustrated random walks}
The calculation of the $B$ matrix elements for frustrated random walks is similar to that for simple random walks. Here, we still have $B_{ij} = T^F_{ij}, \forall i, j \in \mathbb{V} - \{t\}$ and $B_{it} = 0, \forall i \in \mathbb{V} - \{t\}$. The existence of target node $t$ still renders $B$ a non-Markov matrix in that $\sum_{j}B_{ij} = 1$ if $i$ and $t$ do not lie simultaneously in any hyperedge, and $\sum_{j}B_{ij} = 1 - T^F_{it} < 1$ otherwise. This matrix is again a substochastic matrix. However, there are two major differences we want to emphasize. The first is that due to the introduction of the acceptance probability into the frustrated random walks, the transition probability from an \textit{adherent} to the target no longer obeys the sharp $\delta$ distribution, but a geometric distribution, i.e., $P(N_{t}^{(adherent)} = n) = (1-p)^{n-1} p$, where $p$ is the probability of $t$ accepting the transition from the adherent. Consequently, the expected hitting time from the adherent to the target is $1/p$. The second is that the transition from node $i$ to $j$ may be declined, which means $B_{ii} \ne 0$ for at least one $i \in \mathbb{V}$. This guarantees that $B$ is a primitive matrix\cite{meyer2023matrix}, for which the inequality (\ref{spectral_radius_range}) holds. The same proof in the previous subsection guarantees that $\rho(B) < 1$. 

\subsubsection{Numerical computation of expected hitting times}
\label{numerical_method}
Now that we have calculated the elements of the $B$ matrix, we can continue to solve Eq. (\ref{recursive_equation}) as 
\begin{align}
\boldsymbol{X}_{n}  = B^{n-1} \boldsymbol{X}_{1}, n \ge 1. 
\label{iterative_solution}
\end{align}
Here, the vector component $\boldsymbol{X}_{1}^{(i)}$ is the probability of hitting the target starting from node $i$ in the first step, a probability that can be conveniently read out from the transition matrix. By definition, the expected hitting times from an arbitrary node to the target node is 
\begin{align}
\mathbb{E}\boldsymbol{N}_t & = \sum_{n = 1}^{\infty} n P(\boldsymbol{N}_t = n) := \sum_{n = 1}^{\infty} n \boldsymbol{X}_{n} 
\end{align}
Plugging Eq. (\ref{iterative_solution}) into the above definition yields an expression for the expected hitting times that depends both on the initial condition $\boldsymbol{X}_{1}$ and the coefficient matrix $B$, that is, 
\begin{align}
\mathbb{E}\boldsymbol{N}_t =  \sum_{n = 1}^{\infty} n B^{n-1} \boldsymbol{X}_{1} 
\end{align}
The above equation involves an infinite summation and the powers of the $B$ matrix, and is thus difficult to compute numerically. However, we can devise a convenient method to convert the above summation into an equivalent system of linear equations that we can efficiently solve using highly optimized linear algebra packages. To do this, we first define the probability generating function, which is 
\begin{align}
\boldsymbol{f}(z) = \sum_{n = 1}^{\infty} \boldsymbol{X}_{n} z^{n}, |z| \le 1. 
\label{probability_generating_function}
\end{align}
It is straightforward to show that the radius of convergence of the above power series is larger than 1. More precisely, we can pinpoint its radius of convergence at $R = 1/\rho(B)$, where $\rho(B)$ is the spectral radius of the coefficient matrix $B$. A detailed derivation of the radius of convergence of Eq. (\ref{probability_generating_function}) can be found in Appendix \ref{appendix_3}. We have already shown that $\rho(B)$ is guaranteed to be smaller than one for either simple or frustrated random walks.  With the help of Eq. (\ref{iterative_solution}), the probability generating function can be evaluated exactly as 
\begin{align}
\label{probability_generating_function_closed_form}
\boldsymbol{f}(z) &= \sum_{n = 1}^{\infty} B^{n-1} \boldsymbol{X}_{1} z^{n} \\\nonumber 
& = z (I - zB)^{-1} \boldsymbol{X}_1, |z| \le 1. 
\end{align}
Here, the summation can be evaluated in closed form because the radius of convergence of the power series is larger than one. Still due to the fact that radius of convergence is larger than 1, we can differentiate the infinite series in Eq. (\ref{probability_generating_function}) term by term at $z = 1$ and obtain an equivalent expression for the expected hitting times as 
\begin{align}
\boldsymbol{f}^{\prime}(z=1) &= \sum_{n = 1}^{\infty} n B^{n-1} \boldsymbol{X}_{1} z^{n-1}\Big\vert_{z = 1} \\\nonumber
&= \mathbb{E}\boldsymbol{N}_t 
\end{align}
Therefore, once we have the probability generating function, we can easily calculate the expected hitting times. An example of calculating expected hitting times using this method is given in Appendix \ref{appendix_1}. However, when the dimension of $B$ is huge, analytical evaluation of Eq. (\ref{probability_generating_function}) is computationally prohibitive, which compels us to resort to numerical methods. Multiplying both sides of Eq. (\ref{probability_generating_function_closed_form}) with $I - zB$ and taking the first order derivative with respect to $z$ at $z = 1$ yields 
\begin{align}
(I - B) \boldsymbol{f}^{\prime}(z=1) = \boldsymbol{f}(z = 1)
\label{linear_equations}
\end{align}
The coefficient matrix in the above equation is invertible due to the fact that $\rho(B) < 1$. By definition, $\boldsymbol{f}(z = 1)$, which represents the sum of all probabilities, is a column vector each element of which is 1. Therefore, to calculate the expected hitting times $\mathbb{E}\boldsymbol{N}_t = \boldsymbol{f}^{\prime}(z=1)$, we only need to solve the linear equation (\ref{linear_equations}), which is straightforward to accomplish numerically. It is noteworthy that in Eq. (\ref{linear_equations}), the coefficient matrix $B$ depends both on the hypergraph structure and the target node $t$, and thus target information is implicit in that equation. For real-world hypergraph data, the coefficient matrix in Eq. (\ref{linear_equations}) is generally large and sparse. Moreover, under the framework of frustrated random walks, the coefficient matrix $I - B$ is positive definite. The best method to numerically solve linear equations with large, sparse and positive definite coefficient matrix is the conjugate gradient method\cite{hestenes1952methods}. For simple random walks, $I - B$ is no longer positive definite. Fortunately, variants of conjugate gradient method already exist for such non-positive-definite matrices. Given the rapidity and efficacy of conjugate gradient method and its variants in solving large and sparse linear systems, we will use them throughout our paper. 

Our method for calculating hypergraph node distances can be summarized in Algorithm \ref{hitting_time_pseudocode}. This algorithm applies to both simple random walks and frustrated random walks. The only difference lies in the calculation of the probability transition matrix. For sake of simplicity, in Algorithm \ref{hitting_time_pseudocode}, we only show how to calculate the expected hitting times of simple random walks. The extension of this algorithm to frustrated random walks is straightforward. 
\begin{algorithm}[H]
\caption{CalculateExpectedHittingTime(H, t)}
\label{hitting_time_pseudocode}
\begin{flushleft}
\textbf{Input}: hypergraph $H$\\
\hspace*{\algorithmicindent}target node $t$\\
\textbf{Return}: expected hitting times $\boldsymbol{N}_t$
\end{flushleft}
\begin{algorithmic}[1]
\STATE Calculate transition probabilities using Eq. (\ref{srw_transition_matrix_new}). 
\STATE Set node $t$ as the target. 
\STATE Calculate transition matrix $B$ from hypergraph structure and target node $t$, following the procedures in Section \ref{SRW_B_matrix}. 
\STATE $d \leftarrow B.dimension$
\STATE $\mathbbm{1} \leftarrow \text{column vector of all 1's, with shape = (d, 1)}$
\STATE Solve $(I - B) \boldsymbol{x} = \mathbbm{1}$ using a variant of conjugate gradient method. 
\STATE Obtain expected hitting times $\boldsymbol{N}_t \leftarrow  \boldsymbol{x}$
\end{algorithmic}
\end{algorithm}

\section{Experimental results}
In the previous sections, we have outlined two random walk algorithms, the simple random walk (SRW) and frustrated random walk (FRW), and showed that we can compute their expected hitting times in a unified framework. In this section, we will use real-world hypergraph data sets to show that we can use the expected hitting times as node distances to find each node's nearest neighbors. We also compare their results with that of DeepWalk. 

\subsection{Hypergraphs with ground-truth node labels}
In this section, we will test our methods using two hypergraph data sets with ground truth node labels. We construct the first data set from arXiv and obtain the second data set from Ref. \cite{chodrow2021hypergraph}. We note here that both hypergraphs are scale-free and lightly weighted. A hypergraph is scale-free if its node degree distribution follows a power law, and is lightly weighted if its maximum node degree is much smaller than the number of nodes in the hypergraph. 

\begin{figure}[h!]
\centering
\includegraphics[width=0.85\linewidth]{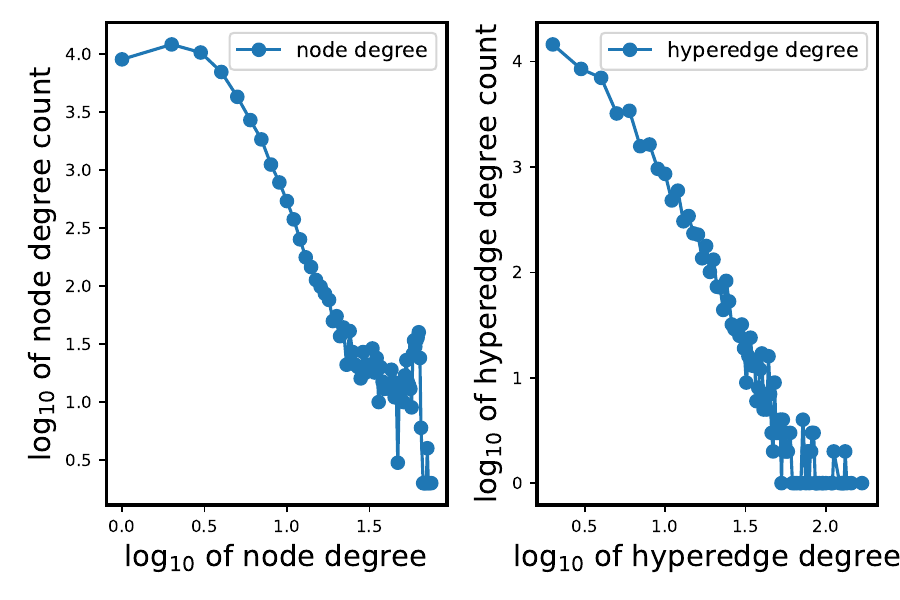}
\caption{$\log_{10}-\log_{10}$ plot of node and hyperedge degree distribution for arXiv data set. Although the node degree distribution does not strictly follows the power law, the power law is pretty obvious for hyperedge degree distribution. }
\label{arxiv_node_hyperedge_degree_log}
\end{figure}

First, we create hypergraphs using articles published in arXiv under the category of physics from 2017 to 2020. Hypergraph nodes represent articles, and articles written by the same author are collected into the same hyperedge. From the hypergraph, we extract its largest connected component, which contains $52,144$ nodes (articles) and $45,188$ hyperedges (authors). The node and hyperedge degree distributions are shown in Fig. \ref{arxiv_node_hyperedge_degree_log}. In this data set, the node (i.e., article) degree represents the number of authors of the article, and the hyperedge (author) degree counts the number of articles written by the author. 

Each arXiv article has one or more subjects, which we can use as its ground truth label. Using the methods described in this paper, we can calculate hypergraph node distances, and thus find each article's nearest neighbors. Intuitively, the shorter the distance between two articles, the more similar they should be to each other. We can quantify two articles' similarity by calculating the Jaccard similarity between their subjects. For comparison, we find each article's nearest neighbors using simple random walks (SRW), frustrated random walks (FRW) and DeepWalk, and calculate the mean Jaccard similarity between the target article and its top ten nearest neighbors. For one specific article, each method gives a mean Jaccard similarity score. The higher the mean Jaccard similarity, the better the method. We randomly select 12 articles, and for each method, calculate these 12 articles' average Jaccard similarity score. The average score for each method is shown in Table \ref{Jaccard_similarity_comparison}. 

\begin{table}
\centering
\begin{tabular}{lccr}
\hline\hline
Method name & SRW & FRW & DeepWalk\\
\hline
Jaccard similarity & 0.4092 & \textbf{0.4615} &  0.4599\\
\hline\hline
\end{tabular}\\
\caption{Comparison of Jaccard similarities from three methods. FRW is on par with DeepWalk, and beats SRW with a large margin. }
\label{Jaccard_similarity_comparison}
\end{table}

From table \ref{Jaccard_similarity_comparison}, we can see that FRW has the highest Jaccard similarity (denoted in bold type in the table). The arXiv hypergraph is approximately scale-free and lightly weighted. While FRW achieves a better result, we will further show in Section \ref{scale_free_heavily_weighted} that the advantage of FRW over SRW is most obvious on heavily weighted scale-free hypergraphs. 

\begin{figure}[h!]
\centering
\includegraphics[width=0.85\linewidth]{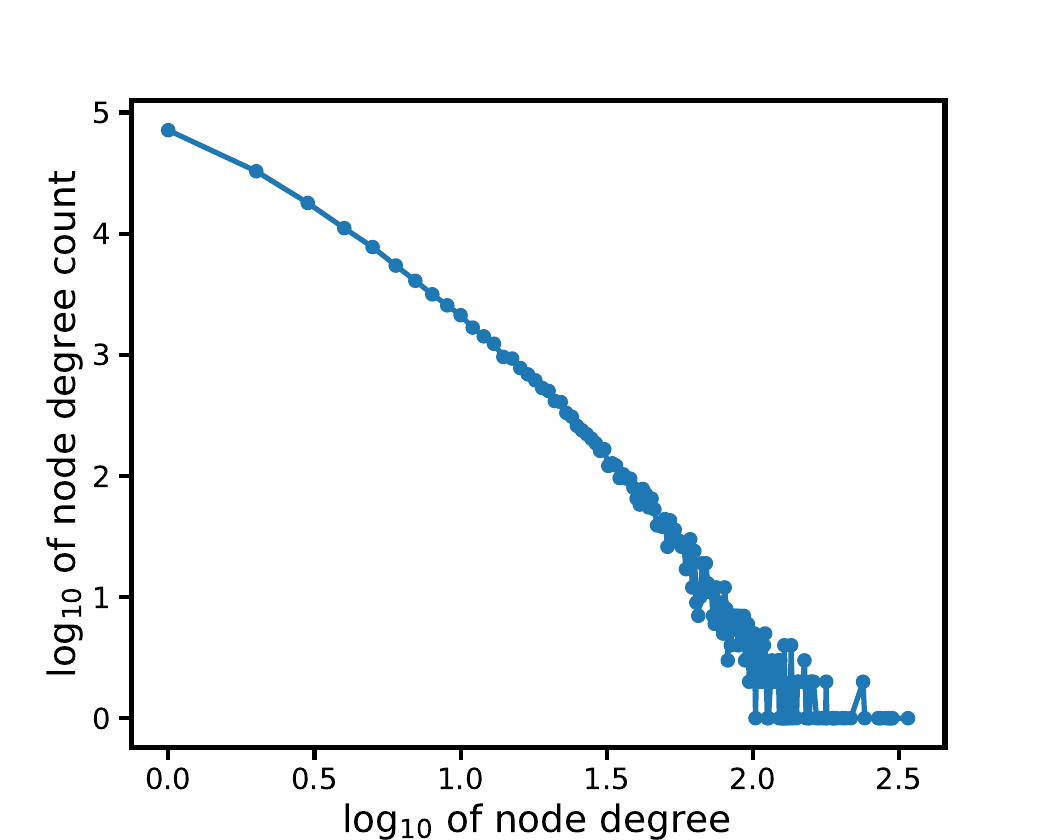}
\caption{$\log_{10}-\log_{10}$ plot of node degree distribution for trivago data set. This data set is a scale-free hypergraph. }
\label{trivago_node_degree}
\end{figure}

The second data set we experiment with is the trivago-clicks data set\cite{trivago_dataset}. This data set has been introduced by the authors of Ref. \cite{chodrow2021hypergraph} as a collection of behavior data from users browsing trivago.com while trying to book a hotel. In this hypergraph data set, nodes represent accommodations (mostly hotels) browsed by users, and hyperedges represent a user's browsing history in the same browsing session before checkout (i.e., before an order is placed). Each node (accommodation) is associated to a country, which can be used as the node's label. The data set contains $172,738$ nodes and $233,202$ hyperedges. The number of distinct node labels (countries or regions) is $160$. The node degree follows a power law, yet the node degrees are generally small, ranging from $1$ to $339$, which is significantly smaller than the hypergraph node number. As a result, this hypergraph is still lightly weighted. A $\log_{10}-\log_{10}$ plot of the node degree distribution is shown in Fig. \ref{trivago_node_degree}. 

 Using SRW, FRW and DeepWalk, we can find each node's nearest neighbors. Generally, we expect a node and its nearest neighbors to be located in the same country (having the same label). We can quantify this similarity by counting the proportion of nearest neighbors sharing the same label as that of the target node. We still randomly select 12 nodes from this hypergraph, and calculate the similarity between a target node and its top 100 nearest neighbors using the three methods. The results are shown in Table \ref{trivago_similarity_comparison}. On this data set, the results from the three methods are almost identical, with more than 97\% nearest neighbors having the same label as that of the target node, indicating the effectiveness of all three methods in detecting nearest neighbors. 

\begin{table}
\centering
\begin{tabular}{lccr}
\hline\hline
Method name & SRW & FRW & DeepWalk\\
\hline
Similarity & 97.08\% & 97.75\% &  \textbf{97.92\%}\\
\hline\hline
\end{tabular}\\
\caption{Comparison of similarities resulting from SRW, FRW and DeepWalk methods using the trivago data set. A larger similarity indicates a better method. In this data set, all three methods give almost equally good results. }
\label{trivago_similarity_comparison}
\end{table}

In this section, we have shown the effectiveness of FRW in measuring node distances on real-world hypergraphs. Although FRW is effective, its advantage compared to SRW is not yet fully visible. In the next section, we will demonstrate the full advantage of FRW using more heavily-weighted and scale-free hypergraphs.

\subsection{Heavily-weighted and scale-free hypergraphs without node labels}
\label{scale_free_heavily_weighted}
As already stated in the previous sections, the node degree distribution of a scale-free hypergraph that is generated according to the principle of preferential attachment follows the power law\cite{barabasi1999emergence}. A hypergraph is called heavily weighted if its node degree range spans several orders of magnitude, and the maximum node degree is at least as large as the hypergraph node number. Since we have designed the frustrated random walks (FRW) with preferential attachment effect in mind, we expect this method to more accurately compute node distances on scale-free hypergraphs, as will be demonstrated in this section. 

Unlike the arXiv hypergraph, the hypergraphs in this section do not have ground truth labels for their nodes, which necessitates us to compare our random walk results with that obtained from DeepWalk. We have chosen DeepWalk as our benchmark for three reasons. First, it is easy to generalize DeepWalk from graphs to hypergraphs, whereas for the other node embedding algorithms, such a generalization is highly non-trivial. Second, among various node embedding algorithms, DeepWalk excels at dealing with scale-free networks since it is essentially a language model, and word frequency in natural languages follows the power law distribution (Zipf's law)\cite{perozzi2014deepwalk}. Third, the two hypergraphs in this section are both derived from novels written in natural languages, a domain where DeepWalk performs particularly well.

\begin{figure}[h!]
\centering
\includegraphics[width=0.85\linewidth]{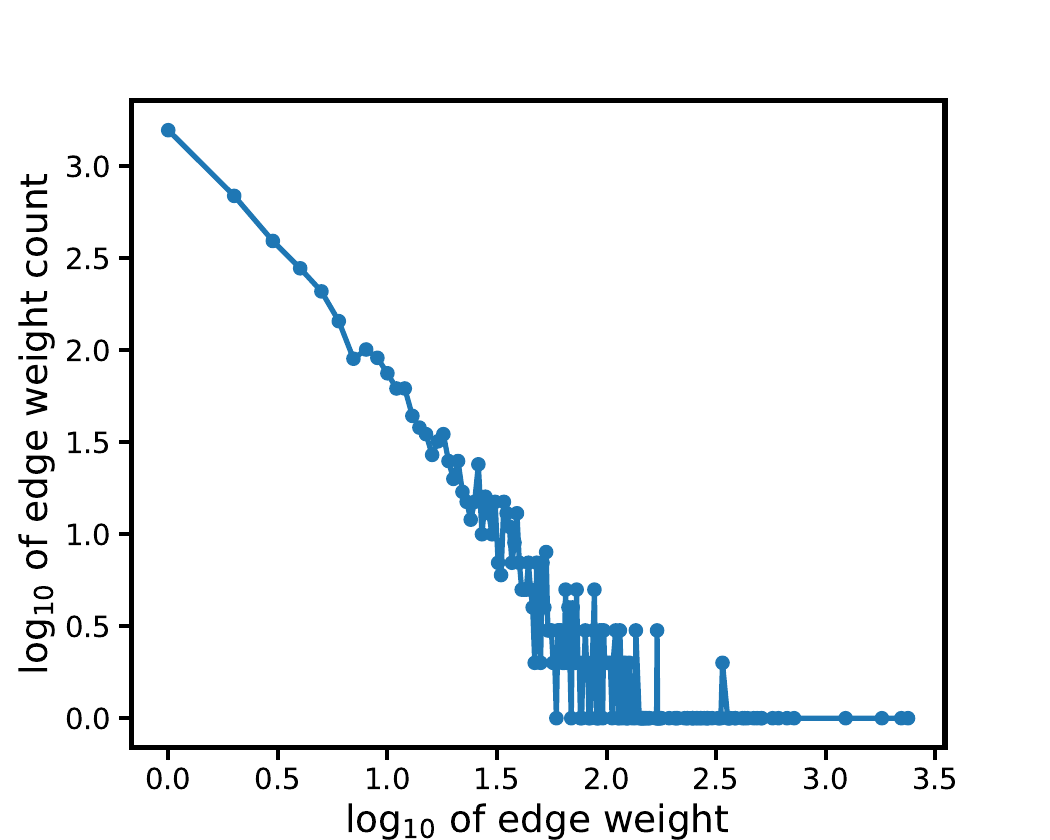}
\caption{$\log_{10}-\log_{10}$ plot of edge weight distribution in \textit{Harry Potter} data set. This figure shows the data set is a heavily-weighted, scale-free hypergraph. }
\label{harry_potter_edge_weight_log}
\end{figure}

We first apply our method to a hypergraph constructed from the \textit{Harry Potter} novels written by J.K. Rowling. This novel was chosen for its intricate plots, complex character relationships and world-wide popularity. In our hypergraph, characters are represented as nodes, and each scene in the novel is represented as a hyperedge. Characters are grouped into the same hyperedge if they appear in the same scene. Node degree indicates the number of times a node (character) appears in the novel. This hypergraph contains $183$ nodes and $3557$ hyperedges. The maximum node degree is 4548, which is much larger than node number. Thus, this hypergraph is heavily-weighted. In this hypergraph, the node number is too small for the power law to be obviously visible. One way to see power law in this data set is to expand the hyperedges into edges, and plot the $\log_{10}-\log_{10}$ distribution of the edge weights, as shown in Fig. \ref{harry_potter_edge_weight_log}.

From our knowledge of the novels, we know that Ron Weasley and Hermione Granger are Harry Potter's best friends. Thus, a good node-distance-computing algorithm should find these two characters as Harry Potter's top two nearest neighbors, meaning their distances to Harry Potter should be the smallest among all nodes. This is the golden criteria we employ in this section to check the validity of our algorithms. We use SRW, FRW, and DeepWalk to rank the nodes by their distances from near to far with respect to Harry Potter, and list our results in Table \ref{harry_potter}. From the table, we can see that both FRW and DeepWalk list Ron Weasley and Hermione Granger as the nearest neighbors to Harry Potter, yet SRW lists Hokey as the nearest neighbor, which is in sharp contrast with our intuition. This corroborates our assertion that FRW is more suitable than SRW for describing real-world human relationships. 

\begin{table}
\centering
\begin{tabular}{lll}
\hline\hline
SRW & FRW & DeepWalk \\
\hline
Hokey & Ron Weasley & Ron Weasley\\
Morfin Gaunt & Hermione Granger & Hermione Granger\\
Merope Gaunt & Severus Snape & Albus Dumbledore\\
Marge Dursley & Sirius Black & Severus Snape\\
Mafalda Hopkirk & Fred Weasley & Ginny Weasley\\
Ignotus Peverell & George Weasley & Minerva McGonagall\\
Helena Ravenclaw & Albus Dumbledore & Voldemort\\
Cole & Rubeus Hagrid & Rubeus Hagrid\\
Mary Cattermole & Ginny Weasley & Fred Weasley\\
Mary Riddle & Draco Malfoy & Sirius Black\\
\hline\hline
\end{tabular}
\caption{Nearest neighbors of Harry Potter (rank order is 1/183) according to different methods. The results from FRW and DeepWalk are consistent with our expectation, meaning  Ron Weasley and Hermione Granger are nearest to Harry Potter, and beat that from SRW. }
\label{harry_potter}
\end{table}

To further compare the results from SRW, FRW and DeepWalk, we list in Table \ref{lucius_neighbors} the nearest neighbors of a peripheral character Lucius Malfoy whose rank order by node degree is 46/183. For such a non-central node, the preferential attachment effect which compromises the performance of SRW is much reduced, thus enabling all three methods to yield results that are consistent with human judgement. 

\begin{table}
\centering
\begin{tabular}{lll}
\hline\hline
SRW & FRW & DeepWalk \\
\hline
Charity Burbage & Narcissa Malfoy & Bellatrix Lestrange\\
Narcissa Malfoy & Bellatrix Lestrange & Charity Burbage\\
Andromeda Tonks & Augustus Rookwood & Augustus Rookwood\\
Rabastan Lestrange & Rodolphus Lestrange & Draco Malfoy\\
Rodolphus Lestrange & Rabastan Lestrange & Narcissa Malfoy\\
Augustus Rookwood & Theodore Nott & Sirius Black\\
\hline\hline
\end{tabular}
\caption{Nearest neighbors of Lucius Malfoy (rank order is 46/183) from different methods. All three methods give results consistent with human intuition. }
\label{lucius_neighbors}
\end{table}

For each target node in the \textit{Harry Potter} hypergraph, we can use SRW, FRW and DeepWalk to rank the other nodes with respect to the target. To systematically compare the ranking results from SRW and FRW, we use DeepWalk results as our benchmark, and calculate the Spearman correlation coefficients between SRW/FRW ranking results and that from DeepWalk. In Table \ref{harry_potter_spearman}, we show the Spearman correlation coefficients for a list of targets. For the targets to be representative, we select nodes with high, medium and low degrees. From the table, we can see that the ranking results of FRW and DeepWalk are mostly consistent with each other for all the nodes, while SRW can yield results consistent with DeepWalk only for peripheral nodes with low degrees. This is understandable since we have already argued in Section \ref{srw_insufficiency} that in SRW, the minions can have unreasonably small distances with respect to the protagonists, which motivates us to introduce the concept of FRW. 

\begin{table*}
\centering
\begin{tabular}{cccccccc}
\hline\hline
Target& Harry Potter & Ron Weasley & Hermione Granger & Voldemort & Gilderoy Lockhart & Michael Corner & Scorpius Malfoy\\
\hline
Node degree & 3317 & 1867 & 1710 & 568 & 77 & 18 & 1\\
SRW & -0.0726 & 0.1266 & 0.0869 & 0.2525 & 0.4957 & \textbf{0.7081}  & \textbf{0.5069} \\
FRW & \textbf{0.5882}  & \textbf{0.6517}  & \textbf{0.5747}  & \textbf{0.5831}  & \textbf{0.5485}  & 0.6407 & 0.4008\\
\hline\hline
\end{tabular}
\caption{Results from the \textit{Harry Potter} data set. Spearman correlation coefficients using DeepWalk results as benchmark, for nodes of high, medium and low degrees. A large coefficient means a high similarity in ranking results with that of DeepWalk. }
\label{harry_potter_spearman}
\end{table*}

\begin{figure}[h!]
\centering
\includegraphics[width=0.85\linewidth]{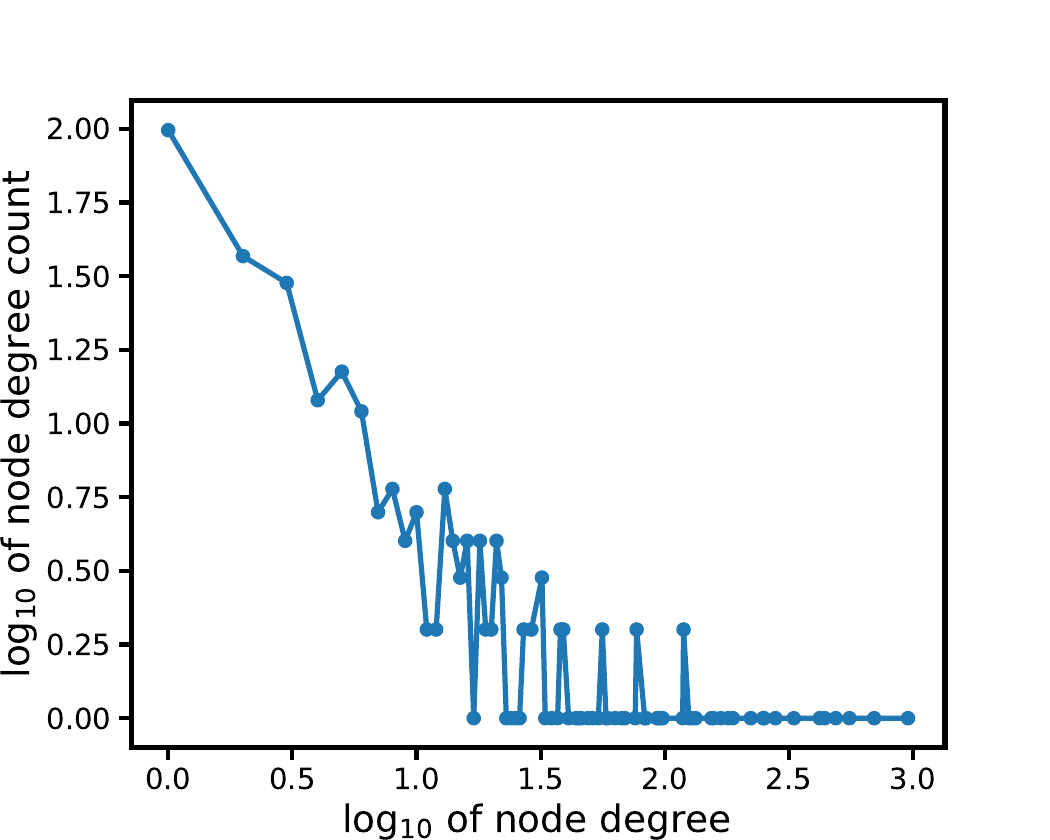}
\caption{$\log_{10}-\log_{10}$ plot of node degree distribution in \textit{Dream of the Red Chamber}. This figure shows the data set is a heavily-weighted, scale-free hypergraph. }
\label{node_degree_log_red_chamber_dream}
\end{figure}

We then apply our methods to a hypergraph created from \textit{Dream of the Red Chamber}, a novel written by Cao Xueqin. This novel is also famous for its complicated plots and tangled character relationships. Still, we represent characters using nodes and characters appearing in the same scene are grouped into the same hyperedge. This hypergraph contains $324$ nodes and $1,524$ hyperedges. The node degree distribution follows a power law, as shown in Fig \ref{node_degree_log_red_chamber_dream}. The node degrees range from $1$ to $954$ which is much larger than the node number, meaning this hypergraph is also heavily weighted. We will show that for such a hypergraph, the FRW method is most suitable for calculating node distances. 

To see this, we will again use DeepWalk as our benchmark, and use SRW and FRW to obtain the nearest neighbors for each target node. The Spearman correlation coefficients between SRW/FRW results and DeepWalk results quantify the accuracy of ranking results from SRW and FRW. In Table \ref{red_chamber_dream_spearman}, we show the Spearman correlation coefficients for nodes of various degrees. From this table, we can reach a similar conclusion as that with \textit{Harry Potter} data set. That is, for central nodes with high degrees, FRW and DeepWalk can yield results that meet our intuition, whereas SRW deviates significantly from either FRW or DeepWalk. Only for peripheral nodes of low degrees can the performance of SRW be on par with that of the other two methods. 

\begin{table*}
\centering
\begin{tabular}{cccccccccc}
\hline\hline
Target& Jia Baoyu & Shi Taijun & Wang Xifeng & Lin Daiyu & Jia Tanchun & Jia Yucun & Liu Xianglian & Wei Ruolan & Pan Youan\\
\hline
Node degree & 954 & 697 & 554 & 488 & 188 & 56 & 13 & 1 & 1\\
SRW & 0.1955 & 0.246 & 0.2037 & 0.4534 & 0.3193 & \textbf{0.6288}  & \textbf{0.4955}  & \textbf{0.6913}  & \textbf{0.5783} \\
FRW & \textbf{0.5985}  & \textbf{0.577}  & \textbf{0.5343}  & \textbf{0.6287}  & \textbf{0.5823}  & 0.6039 & 0.463 & 0.6005 & 0.5686\\
\hline\hline
\end{tabular}
\caption{Spearman correlation coefficients using DeepWalk results as benchmark, for \textit{Dream of the Red Chamber} data set.}
\label{red_chamber_dream_spearman}
\end{table*}

\section{Comparison of running speeds}
In the previous sections, we have shown that to find a node's nearest neighbors using SRW/FRW methods, we only need to construct the transition matrix from the hypergraph structure and the target node, and solve a system of linear equations using conjugate gradient method, the time complexity of which is approximately linear, as we will demonstrate in the next section. However, if we want to accomplish this using DeepWalk, we need to map all the nodes to vectors, calculate the cosine distances between the target node vector and all the other node vectors, and finally rank all the other nodes with respect to the target according to cosine distances. The node mapping could be pretty time consuming, thus rendering the DeepWalk inferior in speed compared to SRW/FRW methods. 

To compare the running speeds of these methods, we test our program on three data sets (arXiv, \textit{Dream of the the Red Chamber}, and \textit{Harry Potter}) using a Linux system with 64 processors. When running DeepWalk, we generate random paths by walking $3200$ steps starting from each node in the hypergraph, and use word2vec\cite{mikolov2013efficient, mikolov2013distributed} method to map each node to a vector of size 128. To accelerate the program, the generation of random paths is fully parallelized. To obtain the running speeds for SRW and FRW methods, we randomly select some nodes as targets, and calculate the mean and standard deviation of the times spent on these nodes. The timing results are shown in Table \ref{running_speeds}. 

\begin{table}
\centering
\begin{tabular}{lccc}
\hline\hline
Data set & arXiv & \textit{DRC} & \textit{Harry Potter}\\
\hline
DeepWalk & 841 & 8.28 &  5.72\\
SRW & $3.79 \pm 0.17$ & $0.012 \pm 0.0002$ & $0.0086 \pm 0.0003$ \\
FRW & $20.12 \pm 0.4$ & $0.026 \pm 0.0005$ & $0.018 \pm 0.012$ \\
\hline\hline
\end{tabular}\\
\caption{Running times of DeepWalk, SRW and FRW on three data sets (\textit{DRC} is short for \textit{Dream of the Red Chamber}). The unit for all numbers is "seconds". For the task of finding one node's nearest neighbors, SRW and FRW can beat DeepWalk in running speeds with a large margin. }
\label{running_speeds}
\end{table}

As can be seen from the table, the running speeds of SRW and FRW are much faster than that of DeepWalk. SRW is even faster than FRW, yet as we have shown in the previous sections, for scale-free and heavily weighted hypergraphs, SRW cannot yield results that are on par with DeepWalk or FRW.

\section{Time complexity of the expected hitting time method}
In Algorithm \ref{hitting_time_pseudocode}, we have outlined our method for calculating expected hitting times. For large hypergraphs, the most time consuming part of the algorithm is driven by finding the numerical solution of equation 
\begin{align}
(I-B)\boldsymbol{x} = \mathbbm{1}
\label{basic_linear_equation}
\end{align}
For large and sparse coefficient matrix $B$, it is best that we solve this equation using conjugate gradient method or its variants. Conjugate gradient method involves a series of iteration cycles, and the most time consuming part of each cycle is the evaluation of $Bv$, where $B$ is the probability transition matrix and $v$ is a dense vector. Thus, the time complexity of our method as described in Algorithm \ref{hitting_time_pseudocode} is $N_{iter}\times O(Bv)$, where $N_{iter}$ is the iteration number in conjugate gradient method and $O(Bv)$ is the time complexity of evaluating $Bv$. We will determine $O(Bv)$ first. 

From Eq. (\ref{srw_transition_matrix_new}) and Eq. (\ref{frustrated_transition_probability}), we can see that each of the summation $\sum_{j}B_{ij}v_{j}$ requires $D_i+1$ operations, where $D_{i}$ is the node degree of vertex $i$. Thus, the total number of operations required for computing $Bv$ is $\sum_{i} (D_i + 1) = 2E + V$, where $E$ is the edge number (the number of edges if we expand all hyperedges in the hypergraph) and $V$ is the node number. We thus have $O(Bv) = 2E + V$. 

To determine $N_{iter}$, we note that conjugate gradient method is guaranteed to give the exact solution to Eq. (\ref{basic_linear_equation}) after exactly $N_B$ iterations, where $N_B \lesssim V$ is the dimension of the coefficient matrix. However, for most cases, conjugate gradient method already gives results of high precision when $N_{iter} \ll N_B$. We can thus consider $N_{iter}$ to be a small constant number compared to $N_B$. 

We can therefore conclude that the time complexity of our algorithm is $O(2E +V )$, with the understanding that $N_{iter}$ is a small constant number. For real-world hypergraphs, we generally have $E\propto V$. In such cases, the time complexity can be simplified to $O(V)$. Thus, the time complexity of our method is approximately linear with respect to the hypergraph size. This linear time complexity for the computation of expected hitting times of random walks is a significant advantage it has over the DeepWalk method.

\section{Conclusion}
In this paper, we proposed the concept of frustrated random walks on hypergraphs as an improvement of simple random walks and developd a unified framework to compute their expected hitting times. By leveraging the transition matrices of both simple and frustrated random walks, we derived an analytical formula for their expected hitting times and employed the conjugate gradient method or its variants for numerical solutions. We suggested using the expected hitting times of random walks as hypergraph node distances and compared the results with those obtained from DeepWalk.

Our findings indicate that for lightly-weighted hypergraphs, all three methods produce high-quality results. However, for heavily-weighted and scale-free hypergraphs, the simple random walk method yields results significantly deviating from human intuition due to the preferential attachment effect. In contrast, frustrated random walks, which takes into account the preferential attachment effect via the introduction of acceptance probability, provide a more accurate description of such highly complex hypergraphs. Although DeepWalk, originating as a language model, is also effective for computing node distances on these hypergraphs, our method surpasses DeepWalk in terms of running speed by a large margin. Finally, we analyzed the time complexity of our random walk method and demonstrated that it is approximately linear for sparse hypergraphs.

\appendix

\section{Computation of expected hitting times on an artificial hypergraph}
\label{appendix_1}
\begin{figure}[h!]
\center
\includegraphics[width=0.75\linewidth]{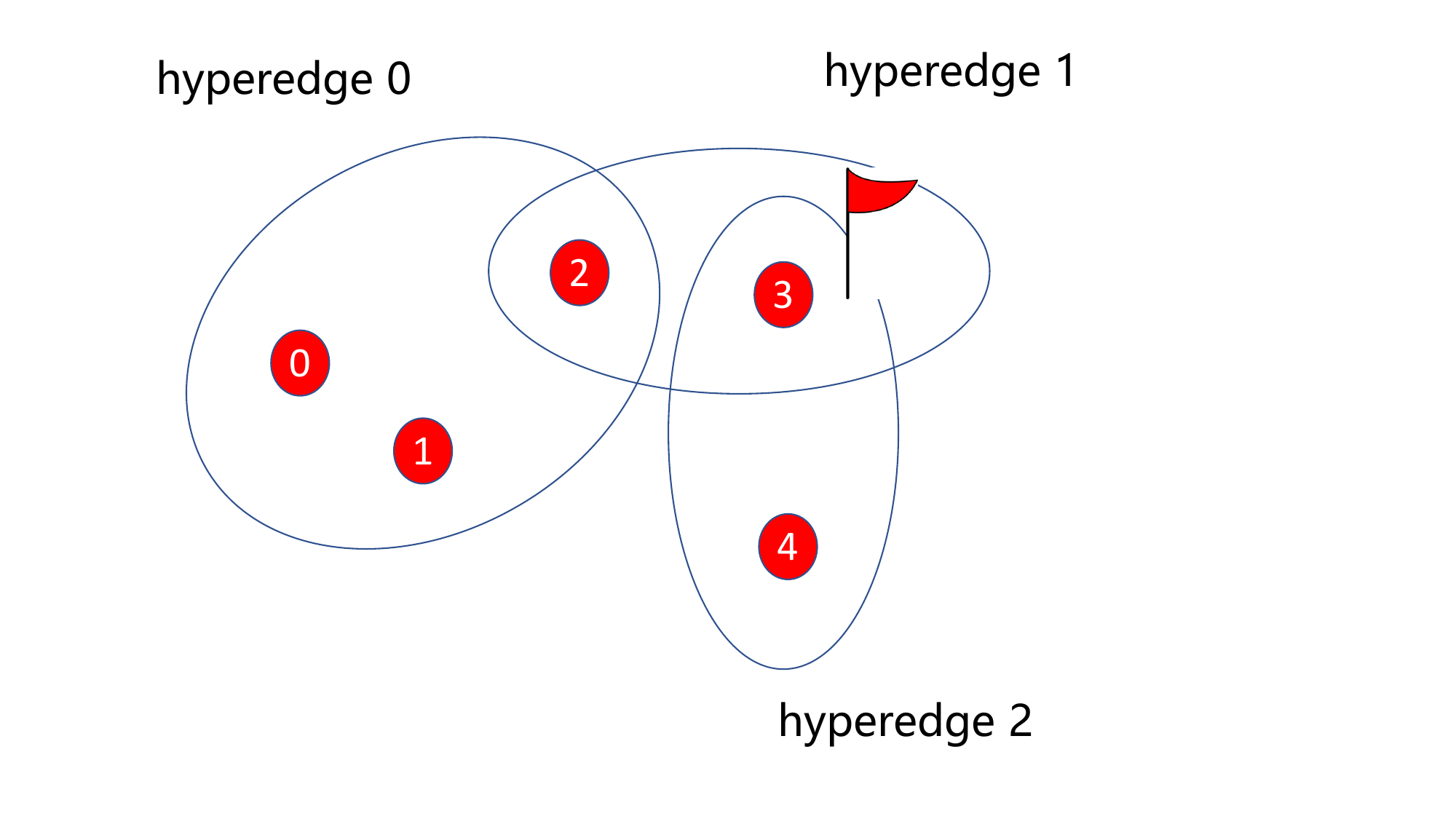}
\caption{An example hypergraph, with node 3 as the target. }
\label{hypergraphexample}
\end{figure}

In this section, we illustrate the procedures of calculating expected hitting times of frustrated random walks on an artificial hypergraph. Monte Carlo simulations are used to validate our numerical results. The expected hitting times of simple random walks can be calculated in a similar manner. We use the hypergraph in Fig. \ref{hypergraphexample} as our example. 

In this hypergraph, there are five nodes and three hyperedges. Its incidence matrix is 
\begin{align}
e = \begin{pmatrix}
1 & 0 & 0  \\
1 & 0 & 0 \\
1 & 1 & 0 \\
0 & 1 & 1 \\
0 & 0 & 1
\end{pmatrix}
\end{align}
We use node 3 as the target node, and calculate the expected hitting times of frustrated random walks starting from nodes \{0, 1, 2, 4\}. First note that, according to our definition in section \ref{unfiedframework}, node 4 is an adherent to target node 3, since node 4 has the target node as its only neighbor. From Eq. (\ref{frustrated_transition_probability}), we can see that the transition probability from node 4 to 3 is $p = \frac{1}{2}$. Thus, the expected hitting time from node 4 to 3 is $\mathbb{E}N_{3}^{(4)} = \frac{1}{p} = 2$. Next we focus on the calculation of expected hitting times starting from nodes \{0, 1, 2\}. 

From Eq. (\ref{frustrated_transition_probability}) and Eq. (\ref{recurrence_equation}), we can establish a system of difference equations for the hitting time probabilities, which is 
\begin{align}
\boldsymbol{X}_{n} = B \boldsymbol{X}_{n-1}, n\ge 2, 
\label{difference_equation}
\end{align}
where the hitting time probability vector is 
\begin{align}
\boldsymbol{X}_{n} = \begin{pmatrix} 
P(N_{3}^{(0)} = n) \\
P(N_{3}^{(1)} = n) \\
P(N_{3}^{(2)} = n)
\end{pmatrix}, 
\end{align}
and the probability transition matrix is 
\begin{align}
B = \begin{pmatrix}
11/20 & 1/4 & 1/5 \\
1/4 & 11/20 & 1/5 \\
1/5 & 1/5 & 1/2
\end{pmatrix}. 
\end{align}

To solve Eq. (\ref{difference_equation}), we need the initial condition which is $\boldsymbol{X}_1 = \begin{pmatrix} 0 & 0 & \frac{1}{10}\end{pmatrix}^{T}$. Using the methods outlined in Section \ref{numerical_method}, we can obtain the probability generating function as 
\begin{align}
\boldsymbol{f}(z) &= z(I - zB)^{-1}\boldsymbol{X}_{1} \\\nonumber
&=  \frac{1}{z(16 z - 65) + 50} \begin{pmatrix}
z^2 \\
z^2 \\
(5 - 4z) z
\end{pmatrix}
\end{align}
First order derivative of the above equation yields the expected hitting times, which are
\begin{align}
\mathbb{E}\boldsymbol{N}_{3} =  \boldsymbol{f}^{\prime}(z  = 1) = \begin{pmatrix}
35 \\
35 \\
30
\end{pmatrix}
\end{align}

We can similarly obtain the above results using Eq. (\ref{linear_equations}), which is 
\begin{align}
\mathbb{E}\boldsymbol{N}_{3} = (I - B)^{-1}\boldsymbol{f}(z=1) = \begin{pmatrix}
35 \\
35 \\
30
\end{pmatrix}
\end{align}

To validate the above results, we write a Monte Carlo program to simulate the random walk process. Running the Monte Carlo simulation 100K times, we obtain the mean value of the hitting times starting from nodes \{0, 1, 2, 4\}. Monte Carlo simulation results and exact results are shown together in Table \ref{hitting_times_table}. 
\begin{table}
\centering
\begin{tabular}{ccc}
\hline\hline
Expected hitting times & Analytical result & Monte Carlo result \\
\hline
$\mathbb{E}N_{3}^{(0)}$ & 35 &  34.91879\\
$\mathbb{E}N_{3}^{(1)}$ & 35 &  35.00548\\
$\mathbb{E}N_{3}^{(2)}$ & 30 &  30.03275 \\
$\mathbb{E}N_{3}^{(4)}$ & 2 & 1.99335 \\
\hline\hline
\end{tabular}\\
\caption{Expected hitting times of frustrated random walks on hypergraph \ref{hypergraphexample} with node 3 as target, starting from nodes \{0, 1, 2, 4\}, using both analytical and Monte Carlo simulation methods. Analytical methods and Monte Carlo simulation yield consistent results, although analytical method can give high-precision results using much shorter computational time. } 
\label{hitting_times_table}
\end{table}
From the table, we can see that exact results are consistent with Monte Carlo simulation results, which justifies our computational method. For large hypergraphs, it is impractical to run Monte Carlo simulations because it is too time consuming. In this case, we will only use numerical methods to compute the expected hitting times. 

\section{Spectral radius of a substochastic matrix}
\label{appendix_2}
In this section, we will prove heuristically that the spectral radius of a substochastic matrix $B$ of the shape $n\times n$ must be smaller than one. A rigorous proof of this theorem can be found in Ref. \cite{666603}. By definition, $B$ must satisfy the following conditions: 
\begin{enumerate}
    \item The matrix is non-negative, meaning that $B_{ij} \ge 0, \forall i, j$. 
    \item $0 < \sum_{j = 1}^{n} B_{ij} \le 1, \forall i$, and $\min_{i}\sum_{j=1}^{n} B_{ij} < 1, \max_{i}\sum_{j=1}^{n} B_{ij}  = 1$. 
    \item $B$ is irreducible. 
\end{enumerate}
When applying matrix $B$ to column vector $\mathbbm{1}$, we obtain a new vector 
\begin{align}
v^{(1)} = B \mathbbm{1}
\end{align}
Since $B$ is substochastic, each element of the new vector $v^{(1)}$ is either 1 or a positive number that is smaller than 1, which we denote as $v^{(1)} < \mathbbm{1}$. Here, for two vectors $\boldsymbol{a}, \boldsymbol{b}$ of the same length, $\boldsymbol{a} < \boldsymbol{b}$ means $a_{i} \le b_{i}, \forall i$ and $a_{i} < b_{i}$ exactly holds for at least one $i$. In this section, we will interpret $``<"$ (and $``>"$) in this manner whenever it appears between two vectors. We thus claim that matrix $B$ describes a random walk that has a sink. When we apply it to a vector each element of which is positive (we will call it a positive vector), we will lose something from the vector, meaning the vector elements will decrease somehow. Applying $B$ consecutively to $\mathbbm{1}$ yields a series of vectors: 
\begin{align}
v^{(k+1)} = B v^{(k)}, \forall k \ge 1
\end{align}
From the substochastic nature of $B$, we have 
\begin{align}
\mathbbm{1} > v^{(1)} > \ldots >  v^{(k)} > v^{(k+1)} > \ldots 
\end{align}
Since the random walk described by $B$ has a sink that devours everything, we eventually have 
\begin{align}
\label{v_infinity}
\lim_{k\rightarrow\infty}v^{(k)} = \lim_{k\rightarrow\infty} B^{k} \mathbbm{1} = 0
\end{align}
That $B$ is non-negative means $B^k$ is also non-negative for any integer $k$. Denoting $B^{\infty} = \lim_{k\rightarrow\infty} B^{k} $, we know from $B^{\infty} \mathbbm{1} = 0$ that 
\begin{align}
B^{\infty} = 0
\end{align}
This means the spectral radius of $B$ is smaller than 1, just as we expected. 

\section{Radius of convergence of the probability generating function}
\label{appendix_3}
In this section, we are going to determine the radius of convergence of the probability generating function defined in Eq. (\ref{probability_generating_function}), which is 
\begin{align}
\boldsymbol{f}(z) &= \sum_{n = 1}^{\infty} \boldsymbol{X}_n z^n \\\nonumber
&= \Big( \sum_{n = 1}^{\infty} B^{n-1}  z^n \Big) \boldsymbol{X}_1
\end{align}

The determination of radius of convergence of function $\boldsymbol{f}(z)$ is tantamount to finding the values of $z$ such that the series $S = \sum_{n=1}^{\infty} B^{n-1} z^{n}$ converges. For frustrated random walks, the coefficient matrix $B$ is real and symmetric, which guarantees that $B$ is diagonalizable and all its eigenvalues are real. In the case of simple random walks, the $B$ matrix is asymmetric and may not be diagonalizable, yet due to the fact that diagonalizable matrices are dense in the set of all matrices, we can still safely assume that $B$ can be diagonalized, although its eigenvalues may be complex. Thus, for an arbitrary transition matrix $B$ in this paper, we can assume with probability one\cite{hetzel2007probability} that there always exists an invertible matrix $U$ and a diagonal matrix $\Lambda$ such that 
\begin{align}
B = U \Lambda U^{-1}
\end{align}
Here, the diagonal matrix $\Lambda$ has all the eigenvalues of $B$ as its diagonal elements, that is, $\Lambda = diag(\lambda_1, \lambda_2, \ldots, \lambda_N)$, where $N$ is the size of $B$ and $\lambda_i, i = 1, 2, \ldots, N$ are its eigenvalues. 
With the knowledge of this, we can rewrite the series $S$ as 
\begin{align}
\label{matrix_series}
S & = \sum_{n = 1}^{\infty} U \Lambda^{n-1} U^{-1} z^n  \\\nonumber
&= U \Big( \sum_{n=1}^{\infty} \Lambda^{n-1} z^n \Big) U^{-1} \\\nonumber
&= U \begin{pmatrix}
    \sum_{n=1}^{\infty} \lambda_1^{n-1} z^n  & & \\
    & \ddots & \\
    & & \sum_{n=1}^{\infty} \lambda_N^{n-1} z^n 
  \end{pmatrix} U^{-1}
\end{align}
For an arbitrary eigenvalue $\lambda_i$ which may be a complex number, in order for the series $\sum_{n=1}^{\infty} \lambda_{i}^{n-1} z^n$ to converge, we must have $|\lambda_i z| < 1$, or equivalently, 
\begin{align}
\label{convergence_radius_one_lambda}
|z| < \frac{1}{|\lambda_i|}
\end{align}
If $\lambda_i = 0$, we shall interpret the above inequality as $|z| < \infty$. 

The inequality (\ref{convergence_radius_one_lambda}) should hold for any of the eigenvalues of $B$, which means  
\begin{align}
|z| < \frac{1}{\max_{1 \le i \le N} |\lambda_i| } = \frac{1}{\rho(B)}
\end{align}
We have thus pinpointed the radius of convergence of the probability generating function at $\frac{1}{\rho(B)} > 1$. 

\bibliography{ref}
\bibliographystyle{apsrev}

\end{document}